\documentclass[letterpaper,twocolumn,10pt]{article}
\usepackage{style}

\usepackage{tabularx}
\usepackage{enumitem}
\usepackage{subfigure}
\usepackage{xspace}
\usepackage{graphicx,lipsum,multicol,caption}
\usepackage{multirow}
\usepackage{wrapfig}

\usepackage{xspace}
\usepackage{xcolor} 
\usepackage{soul} 
\usepackage{balance}

\newif\ifcomment
\commenttrue

\newif\ifdiff
\difftrue

\ifdiff
    \newcommand{\del}[1]{{{\color{red}\st{#1}}}} 
     
\else
    \newcommand\del[1]{}
    
\fi

\ifcomment
    \newcounter{MVNumberOfComments}
    \stepcounter{MVNumberOfComments}
    \newcommand{\mvnote}[1]{\textcolor{blue!50!black}{\small \bf [MV\#\arabic{MVNumberOfComments}\stepcounter{MVNumberOfComments}: #1]}}
\else
    \newcommand\mvnote[1]{}
\fi

\ifcomment
    \newcounter{YZNumberOfComments}
    \stepcounter{YZNumberOfComments}
    \newcommand{\yznote}[1]{\textcolor{orange!90!black}{\small \bf [YZ\#\arabic{YZNumberOfComments}\stepcounter{YZNumberOfComments}: #1]}}
\else
    \newcommand\yznote[1]{}
\fi

\ifcomment
    \newcounter{APNumberOfComments}
    \stepcounter{APNumberOfComments}
    \newcommand{\apnote}[1]{\textcolor{magenta!50!black}{\small \bf [AP\#\arabic{APNumberOfComments}\stepcounter{APNumberOfComments}: #1]}}
\else
    \newcommand\apnote[1]{}
\fi

\ifcomment
    \newcounter{RANumberOfComments}
    \stepcounter{RANumberOfComments}
    \newcommand{\ranote}[1]{\textcolor{cyan!50!black}{\small \bf [RA\#\arabic{APNumberOfComments}\stepcounter{RANumberOfComments}: #1]}}
\else
    \newcommand\ranote[1]{}
\fi

\newcommand{\tool}{{SONIC}\xspace}

\newcommand{\eg}{{e.g.,}\xspace}
\newcommand{\ie}{{\it i.e.,}\xspace}

\begin{document}

\date{}

\title{\Large \bf \tool: Cost-Effective Web Access for Developing Countries}

\author{
Ayush Pandey$^{\dagger}$, Rohail Asim$^{\dagger}$, Jean Louis K. E. Fendji$^{\ddagger}$, Talal Rahwan$^{\dagger}$, Matteo Varvello$^{\mathsection}$, Yasir Zaki$^{\dagger}$ \\
$^{\dagger}$New York University Abu Dhabi, 
$^{\ddagger}$University of Ngaoundere, 
$^{\mathsection}$Nokia Bell Labs
}

\maketitle

\begin{abstract}
    Over 2.6 billion people remain without access to the Internet in 2025. This phenomenon is especially pronounced in developing regions, where cost and infrastructure limitations are major barriers to connectivity. In response, we design \tool, a low-cost, scalable data delivery system that builds on existing infrastructures: FM radio for downlink broadcasting, and SMS for personalized uplink. \tool is motivated by the widespread availability of FM radio and SMS infrastructure in developing regions, along with embedded FM radio tuners in affordable mobile phones. \tool offers several innovations to effectively transmit Web content over sound over FM radio, in a reliable and compressed form. For example, we transmit pre-rendered webpages and leverage pixel interpolation to recover errors at the receiver. We further modify Android to offer a simpler deployment pipeline, supporting a wide range of devices. We deployed \tool at an FM radio station in Cameroon for six weeks with 30 participants. Our results demonstrate a sustained downlink throughput of 10~kbps, less than 20\% loss for a majority of transmissions with signal strength above -90~dbM, and a strong user engagement across both Web browsing and ChatGPT interactions.
\end{abstract}

\begin{figure*}[t]
   \centering
    \subfigure[RTT across cloud service providers.]{
      \begin{minipage}[b]{0.3\textwidth}
         \includegraphics[width=2.1in]{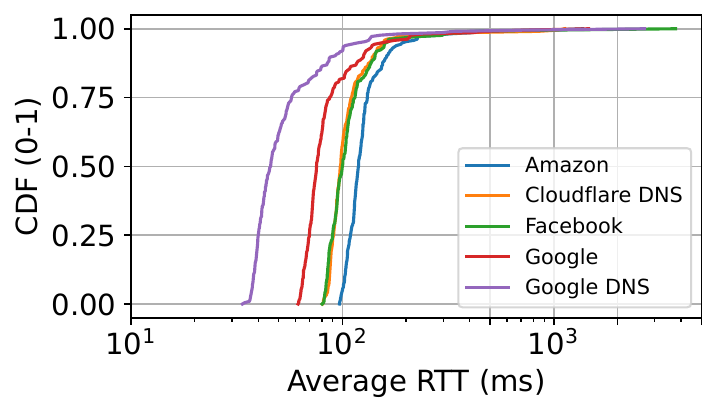}
         \label{fig:rtt}
         \vspace{-10pt}
      \end{minipage}
   }
   \subfigure[Speedtest metrics (latency and bandwidth).]{
      \begin{minipage}[b]{0.3\textwidth}
         \includegraphics[width=2.6in]{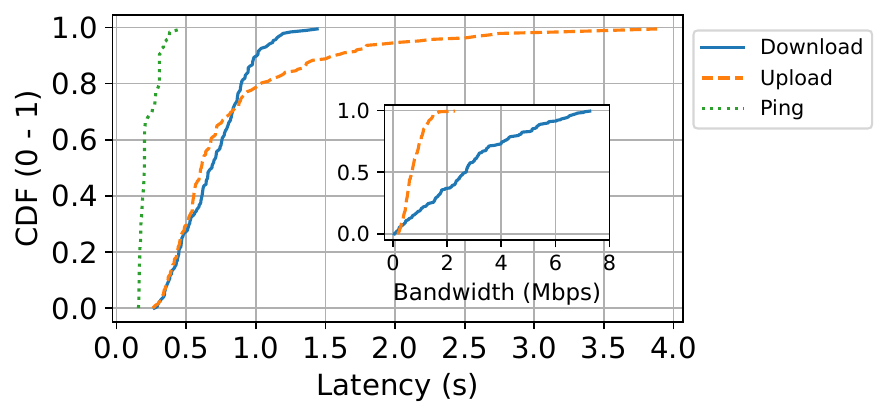}
         \label{fig:latency-bandwidth}
         \vspace{-10pt}
      \end{minipage}
   }\quad\quad\quad
   \subfigure[Web performance metrics (SI, FCP, LCP).]{
      \begin{minipage}[b]{0.3\textwidth}
         \includegraphics[width=2.1in]{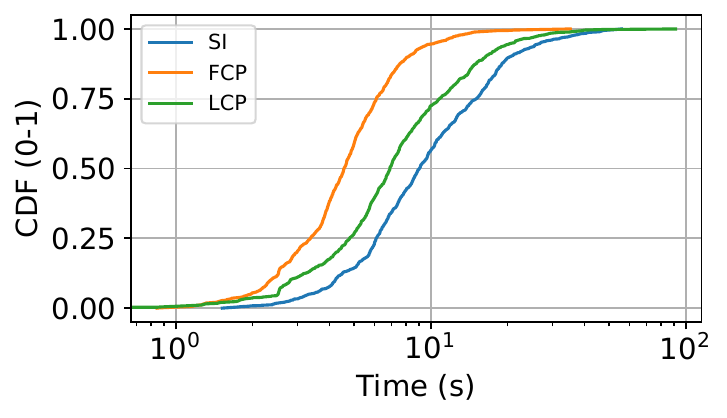}
         \label{fig:page_quality_benchmarking}
         \vspace{-10pt}
      \end{minipage}
   }
   \caption{Network performance of MTN, the best available mobile network provider at our deployment site in Cameroon.}
   \label{fig:sonic-net-tests}
\end{figure*}

\section{Introduction}
The internet has fundamentally reshaped societies worldwide, driving economic growth, fostering new industries, and becoming indispensable in education, healthcare, and employment. Yet, despite its profound influence, 32.4\%---over 2.6 billion people---remain disconnected in 2025~\cite{we_are_social_meltwater_2025}. This \textit{digital divide} is not only due to a lack of infrastructure. According to GSMA Intelligence~\cite{gsma_state_of_mobile_internet_connectivity_2024}, 3.1 billion people live within the coverage area of a mobile broadband network but remain disconnected because they cannot afford the necessary devices and data plans to get online.

The offline population is especially concentrated in developing regions. For example, India has 651 million people (44.7\%) offline, Pakistan 137 million (54.3\%), and Nigeria 128 million (54.6\%)~\cite{we_are_social_meltwater_2025}. In the Central African Republic and South Sudan, over 84\% of the population remains without internet access~\cite{we_are_social_meltwater_2025}. The consequences of being ``unconnected'' go beyond missed economic opportunities; it severely limits access to essential services such as education, healthcare, and financial resources, further exacerbating existing inequalities.

Several initiatives have attempted to bridge this gap. Starlink~\cite{starlink} provides global coverage through low earth orbit (LEO) satellites. Google's Project Loon~\cite{googleLoon} sought to deliver internet via high-altitude balloons, while Facebook's Aquila~\cite{facebookAquila} aimed to use solar-powered drones (both ultimately discontinued due to maintenance costs and scalability challenges). Project Taara~\cite{taara2025} is a recent initiative that transmits data using laser beams over long distances (20~km) at high speeds (20~Gbps). Unfortunately, these efforts remain prohibitively expensive in most developing regions. For example, in Zambia, classified as one of the UN's least developed countries, Starlink's \$40 monthly subscription (plus \$180 hardware fee) is prohibitively expensive compared to the country's \$108 monthly GNI per capita.

In this paper, we build on the preliminary work by~\cite{sonic_conext} which introduced an initial framework to transmit simplified webpages over FM radio. Webpages are pre-rendered as images which are then modulated into audio signals and transmitted via FM radio. We advance~\cite{sonic_conext} as follows: 

\vspace{5pt}
\noindent\textbf{System Architecture.} We design the full software architecture of the \tool server and Android client, which includes modules for content rendering, prefetching, encoding, transmission, decoding, and error correction.

\vspace{5pt}
\noindent\textbf{FM Tuning on Android.} We leverage the internal FM tuner found in many Android smartphones to receive and decode \tool transmissions. We do this by modifying LineageOS, an open-source Android operating system, to enable programmatic access to the FM hardware, allowing other apps to control FM chip tuning and access the received audio stream directly, without needing to root the device.

\vspace{5pt}
\noindent\textbf{LLM Support.} We extend \tool's functionality to support LLM interactions over FM radio. We use the same audio modulation pipeline to allow users to query models like ChatGPT via SMS and receive coherent responses without internet access. We show that LLM responses, due to their smaller size, can be transmitted significantly faster than full webpages.

\vspace{5pt}
\noindent\textbf{Real-world Deployment.} We report on a six-week deployment at a live FM radio station in Cameroon. During this period, 30 participants used the \tool app to request webpages and interact with ChatGPT via SMS. Our evaluation shows that \tool can sustain a transmission rate of 10 kbps, with stable reception achievable at a Received Signal Strength Indicator (RSSI) value up to -90~dBM. Mean decoding accuracy remained at 71\% under real-world conditions.

\section{Motivation}
\label{sec:motivation}
\noindent\textbf{Affordability and QoE in Rural Areas}. Mobile internet adoption is rapidly increasing in low- and middle-income countries~\cite{gsma_state_of_mobile_internet_connectivity_2024}, yet ensuring affordability and a good quality of experience (QoE) remains a challenge. In many remote areas, even when mobile broadband is available, data costs are prohibitively high relative to average income levels. To assess affordability and QoE in a rural setting, we conduct network performance measurements and a survey at the \tool deployment site in Cameroon. 

\begin{figure}[t]
    \centering
    \includegraphics[width=.8\linewidth, clip]{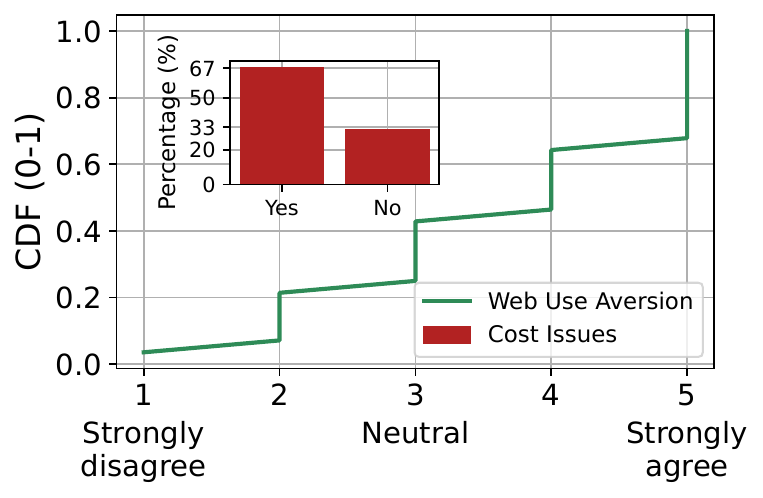}
    \caption{Barriers to Web access from speed and cost. The main plot shows CDF of how \tool users agree with avoiding websites due to slow internet. The inset shows the \% of users indicating that  their browsing is restricted by data costs.}
    \label{fig:survey-internet-experience}
\end{figure}

We select MTN~\cite{mtncameroon}, the best available mobile network provider in the area, and purchase its monthly data plan, which offers 9.2~GB for \$14. Following the methodology in~\cite{amigos}, we use \texttt{mtr}~\cite{mtr}, a tool for measuring latency and tracing network paths to examine routing changes and packet losses. Additionally, we employ \texttt{Speedtest CLI}~\cite{speedtest_cli}, a command-line tool for measuring latency, download, and upload speeds. We also evaluate web performance 
using \texttt{Google Lighthouse}~\cite{lighthouse}, an automated tool that provides key website performance metrics, including Speed Index (SI), First Contentful Paint (FCP), and Largest Contentful Paint (LCP). Given the data limits of our mobile internet plan, we conduct \texttt{mtr} and \texttt{Speedtest} measurements once every three hours throughout our deployment. For Lighthouse evaluations, we rely on data collected in the wild as a response to real \tool users (see Table~\ref{table:deployment-summary}).

Figure~\ref{fig:sonic-net-tests}(a) shows the cumulative distribution function (CDF) of average round-trip time (RTT) towards popular content providers (Amazon, Facebook, and Google), and DNS operators (Cloudflare and Google). The figure shows RTTs higher than 100ms for most measurements and providers, with the exception of Google. As observed in~\cite{varvello2022performance}, this is due to Google footprint in Africa which was also confirmed by our path analysis.

Next, Figure~\ref{fig:sonic-net-tests}(b) summarizes the speedtest analysis. With respect to the RTT to OOKLA servers (ping), the figure shows a similar trend as Figure~\ref{fig:sonic-net-tests}(a). The figure further shows RTT under load, \ie while measuring both download and upload speeds, showing a 3.4x growth (from 200 to 680~ms, at the median) thus suggesting large buffer in use (a phenomenon typically called bufferbloat~\cite{gettys2012bufferbloat}). Despite these large buffers, users experience a median download speed of 2.6~Mbps and an upload speed of 0.69~Mbps---both drastically below the global average---at a mobile internet price comparable to developed nations (\$1.5 per GB)~\cite{amigos}. At these speeds, web performance is extremely affected, as visualized in Figure~\ref{fig:sonic-net-tests}(c). The median SI is at 9~seconds, FCP at 4.5~seconds and LCP at 6.9~seconds, significantly slower than the web vitals threshold (LCP < 2.5~seconds) recommended by Google~\cite{web_vitals}.

The combination of high latencies and low download speeds can frustrate users and discourage regular internet use. Figure~\ref{fig:survey-internet-experience} presents responses to two survey questions assessing barriers to web access at our deployment site in Cameroon. Survey participants are 30 Cameroonians who participated in \tool deployment. The main plot shows the CDF of web use aversion, based on responses to the question: \textit{``To what extent do you agree with the following statement: `I occasionally avoid visiting certain websites because my Internet is too slow to load them.''}' (1 = Strongly disagree, 5 = Strongly agree). Over 75\% of the participants indicate either neutrality or agreement with avoiding websites due to slow internet speeds. The inset plot shows responses to the binary question: \textit{``Is your web browsing experience often restricted by data costs?''} where 67\% of users answered ``Yes.''

These results sheds light on the affordability of web access in low-income regions like Cameroon. Beyond slow speeds, high data costs severely limit both how often and how effectively users can engage with the internet, restricting access to information that many take for granted. Indeed, we also asked participants if they had heard of ChatGPT prior to participating in the experiment. 80\% reported ``No'', revealing a broader lack of exposure to transformative (and popular) technologies simply because the current infrastructure never allowed these tools to reach them.
\begin{figure}[t]
    \centering
    \includegraphics[width=1\linewidth, clip]{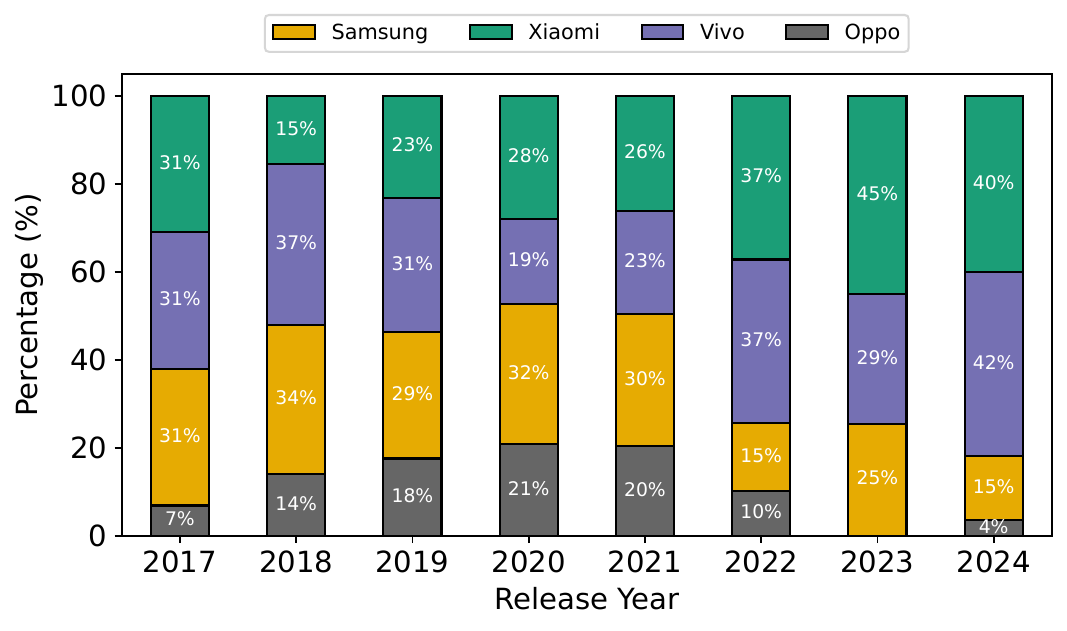}
    \caption{Mobile phones supporting FM receivers for the top four Android brands grouped by release year (2017-2024).}
    \label{fig:mobile-phones-with-fm}
\end{figure}

\begin{figure}[t]
    \centering
    \includegraphics[width=.8\linewidth, clip]{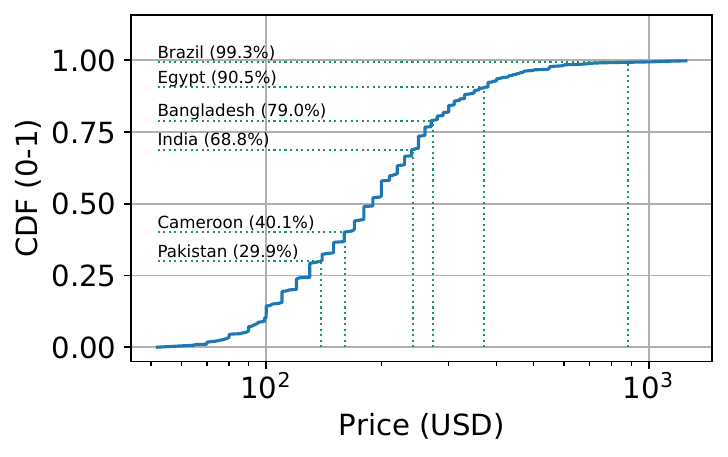}
    \caption{CDF of prices for android phones equipped with an FM radio receiver. Green dotted lines indicate the monthly Gross National Income (GNI) per capita for selected low- and middle-income countries.}
    \label{fig:mobile-phones-with-fm-affordability}
\end{figure}
\vspace{5pt}
\noindent\textbf{FM Radio Availability}. FM radio still remains widely used in developing regions. In a study covering 39 countries in Africa, about 65\% of adults reported listening to radio at least a few times per week, with no major difference in the rural-urban gap in radio access~\cite{afrobarometer2024}. A recent case study in northern Ghana found FM radio to be the most reliable and trusted source of developmental information in rural communities, providing vital content on agriculture, education, and health in local languages~\cite{Antwi-Boateng31122023}. 

Technical studies confirm that FM signals reliably cover large areas, though terrain can influence signal quality. In Nigeria, measurements around a 20~kW FM station showed stable reception up to 50~km, beyond which quality declined due to sandy or silty soil and obstructed line-of-sight~\cite{nigeria2019}. A study from Nepal emphasized the importance of antenna height and placement in extending coverage, even in hilly regions with shadow zones between elevations~\cite{nepal2017}. Despite geographic challenges, FM radio continues to provide consistent, low-cost coverage in most rural areas, with signal strength sufficient for everyday use.

Building on the widespread availability of FM radio connectivity, we next investigate the landscape of FM radio support in mobile phones. Specifically, we analyzed the prevalence of FM receivers in Android smartphones currently available on the market. To do this, we scraped the database of mobile phone specifications from GSMArena~\cite{GSMArena}, identifying which models support FM radio and recording their release years and market prices. We then cross-referenced this information with global economic statistics~\cite{worlddata2025} to assess the practical reach of FM-based data reception.

Figure~\ref{fig:mobile-phones-with-fm} shows the distribution of FM-capable phones released by the top four Android smartphone brands~\cite{counterpoint2025smartphone} over the past eight years. Despite a gradual decline in newer models, FM radio functionality remains prevalent—especially among Xiaomi and Vivo devices, where approximately 40\% of current models still include FM support. In total, these top four brands alone account for 571 FM-capable models, highlighting a significant and readily accessible user base for FM-based data delivery.

We next evaluate the affordability of these 571 FM-capable phone models in low-income countries, using monthly gross national income (GNI) per capita as a benchmark~\cite{worlddata2025}. Figure~\ref{fig:mobile-phones-with-fm-affordability} shows the price distribution of these phones (blue line) alongside GNI thresholds (green dotted lines) for six developing countries: Pakistan, Cameroon, India, Bangladesh, Egypt, and Brazil. The figure shows that a substantial share of devices are within reach of average consumers, from a minimum of 30\% in Pakistan up to 99.3\% in Brazil. These results highlight strong market availability of FM-enabled phones at accessible price points in low and middle-income regions.
\section{\tool}

\begin{figure}[t]
    \centering
    \includegraphics[width=1\linewidth]{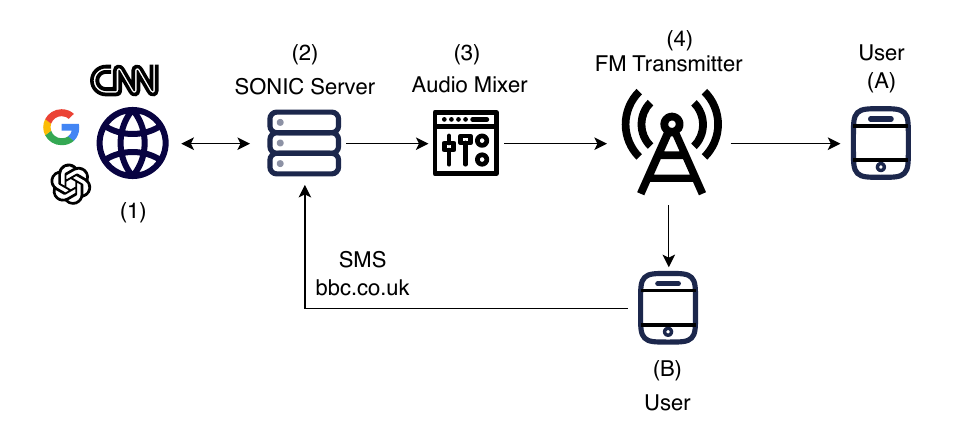}
    \vspace{-0.2in}
    \caption{\tool workflow.}
    \label{fig:workflow}
\end{figure}

Figure \ref{fig:workflow} shows how \tool operates. Users A and B both have the \tool app installed on their smartphones, equipped with chipsets featuring FM receivers. However, only User B's device supports SMS. The \tool app runs on both smartphones with FM receivers tuned to a specific frequency to decode incoming data-over-audio streams.

User B requests a webpage using the \tool app, which sends an SMS containing the desired URL (\eg \texttt{bbc.co.uk}) 
to the \tool server. The server listens for incoming SMS messages, retrieves the webpage, captures a screenshot of the rendered page, and compresses it into a WebP image. The image is then encoded into sound and broadcasted via an FM transmitter. Both User A and User B receive the transmitted webpage on their smartphones. In the following, we describe each \tool component which enables the above workflow.

\subsection{\tool Server}
\begin{figure}[t]
    \centering
    \includegraphics[width=1\linewidth]{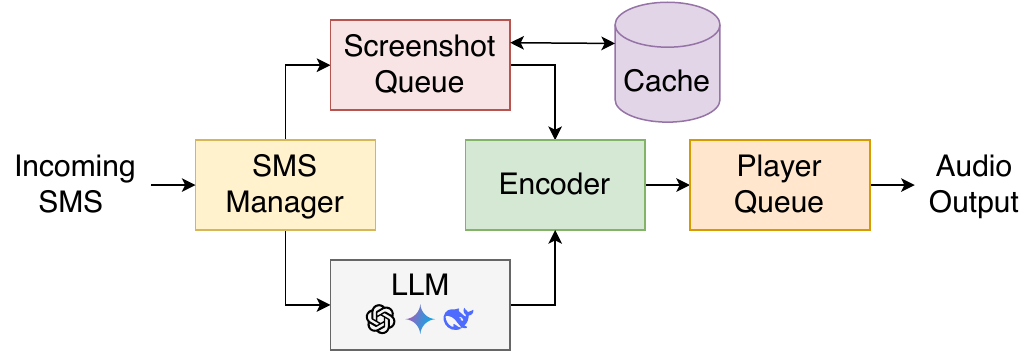}
    \caption{\tool server architecture.}
    \label{fig:sonic-server}
    \vspace{-0.2in}
\end{figure}

Figure~\ref{fig:sonic-server} shows the architecture of the \tool server. The server runs as a Docker~\cite{docker} container on a computer located at a radio station. It consists of several key components that work together to process user requests. The SMS Manager handles incoming messages, while the Screenshot Queue processes webpage URLs in a First Come, First Serve (FCFS) manner. A Cache stores recently requested URLs to avoid redundant processing. The Encoder converts responses into \tool \textit{file format} (see Section~\ref{sec:data-encoding}) and encodes them to audio. Finally, the Player Queue manages the order in which the encoded audio files are played, also following the FCFS policy.

The SMS Manager continuously listens for SMS messages using a USB mobile 
dongle. These messages are received on a phone number assigned to the SIM card inserted into the dongle. Messages sent by the \tool app contain the sender's information in the headers and a payload formatted as: \texttt{<type> <body>} (\eg \texttt{url https://nytimes.com}). When a new message is received, the server classifies it as either an LLM prompt or a webpage URL depending on the \texttt{<type>} identifier.

If the message contains a URL, it is added to the Screenshot Queue. The system checks whether this URL has been retrieved within the current transmission window using the Cache. If not, the Screenshot Queue utilizes Selenium~\cite{selenium} to load the page in Google Chrome with a mobile resolution of an iPhone SE device (375 $\times$ 667 pixels). Once the webpage is fully loaded, a full-page screenshot is taken and resized to a width of 320~pixels. We empirically selected 320~pixels as a sweet spot where both the content layout and text remain comfortably readable to the human eye. The screenshot is then encoded into a \tool file format and subsequently converted into audio using the \texttt{Quiet}~\cite{quiet} library at the Encoder. The resulting audio is then added to the Player Queue. In case of cached requests, the cached response directly moves to the Player Queue.

If the message contains an LLM prompt, the server makes an LLM inference API call to either a locally running LLM, or a cloud API (\eg \texttt{OpenAI Chat Completions API}~\cite{openai_chat_completions}). The response from the LLM is encoded into the \tool file format containing structured metadata and payload. Finally, the server generates a corresponding audio using the \texttt{Quiet} library, which is then added to the Player Queue.

\subsubsection{Encoding}
\label{sec:data-encoding}
Encoding takes place in the Encoder, which consists of two steps: 1) encoding responses to a new file format, referred to as a \textit{sonic~file}; and 2) converting this sonic file into a waveform audio file (WAV). The resulting audio file is then added to the Player Queue.

\begin{figure}[t]
    \center
    \includegraphics[width = 1\linewidth]{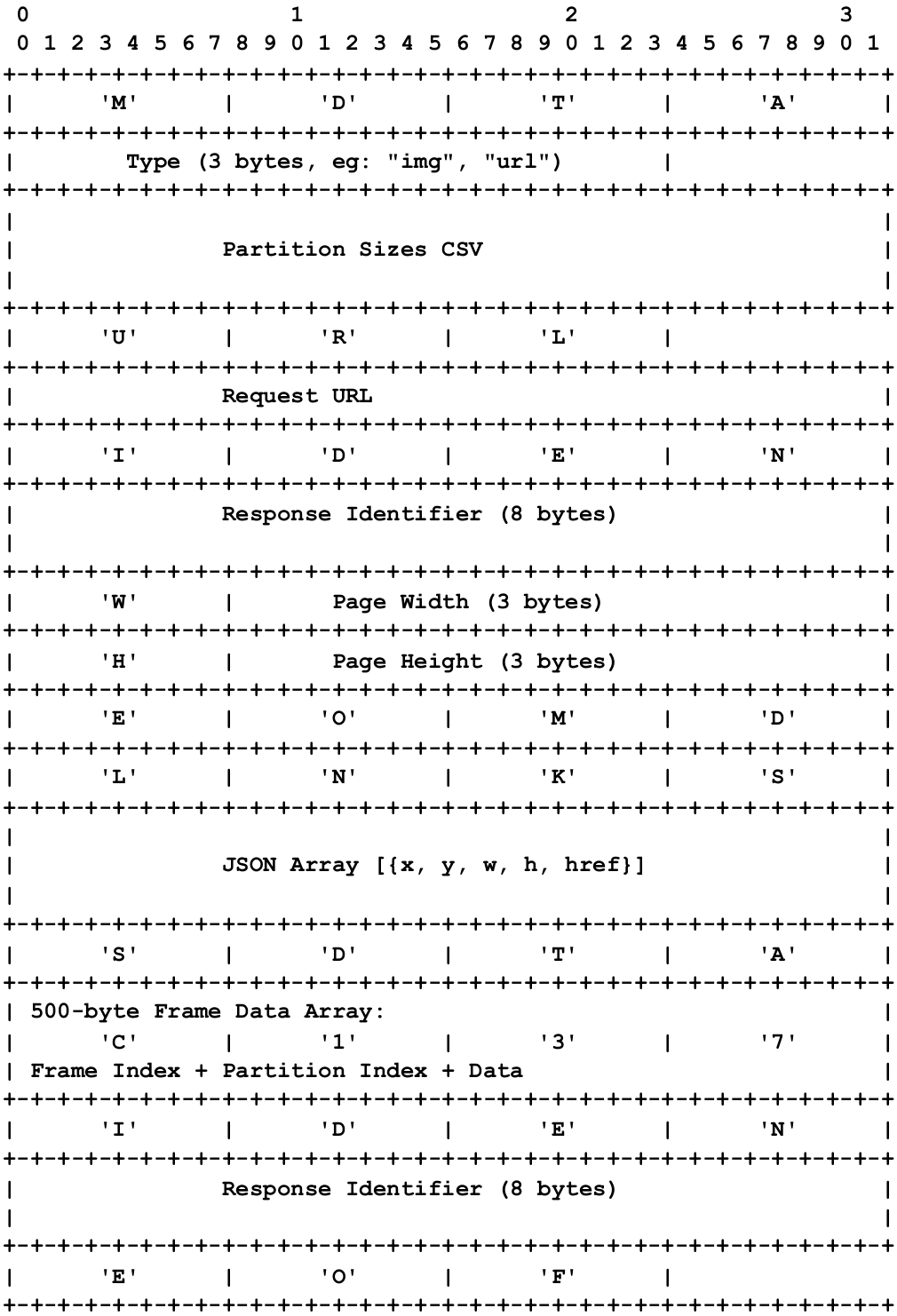}
    \vspace{-0.2in}
    \caption{\tool file format.}
    \label{fig:sonic-file-format}
\end{figure}

\vspace{5pt}
\noindent\textbf{\tool File Format.} \tool uses a new file format that allows the decoder to easily distinguish specific parts of the transmission in absence of a continuous uplink. As illustrated in Figure~\ref{fig:sonic-file-format}, this format includes intermediate headers (such as ``MDTA,'' ``LNKS,'' ``SDTA,'' etc.) that separate internal sections within the metadata and payload. This structure helps reconstruct content in cases when metadata is fully received but only parts of the payload are received properly. Additionally, each payload frame is prefixed with ``C137,'' inspired by \textit{Rick and Morty}'s C-137 dimension~\cite{rickc137}, which helps distinguish the start point of each frame. Furthermore, ``C137'' serves as a keepalive message, allowing the app to notify users that the server is online when transmissions are in progress.

\vspace{5pt}
\noindent\textbf{Webpage Compression.} Unlike encoding LLM interactions, where the payload is simply appended to metadata, encoding webpages presents a significant challenge due to the limited data rates achievable over audio transmission---typically only tens of kbps (see Section~\ref{sec:related}). According to~\cite{webAlmanac}, the average mobile webpage size is approximately 2~MB. Broadcasting such a page via \tool could take tens of minutes. Moreover, devices experiencing poor RSSI may struggle to reconstruct the page, as critical web components, such as JavaScript, may fail in the presence of unrecoverable errors.

To address these challenges, we must: 1) significantly compress webpages, and 2) ensure resilience against noise. Various methods exist for reducing webpage sizes, such as compression proxies~\cite{amp, webLight}, reader modes~\cite{ghasemisharif2019speedreader, safari}, JavaScript cleaners~\cite{JSCleaner, slimweb, jsanalyzer}, and redundant code removal~\cite{kupoluyi2022muzeel, malavolta2023javascript}. These approaches remain vulnerable to noise and require extensive forward error correction (FEC), necessitating a system design that accounts for the worst-case receiver conditions.

Instead, we utilize a solution where performance degrades gracefully as a function of the receiver's RSSI, analogous to how audio quality deteriorates under poor reception. Inspired by~\cite{bozomitu2023drivers, bozomitu2022robust}, which demonstrate how image quality over RDS degrades with RSSI, we opt to transmit images of rendered webpages rather than raw web files (HTML/JavaScript/CSS). This approach provides both compression and resilience: a 2~MB webpage can be compressed into a few hundred KB, and images remain interpretable even if some pixels are lost.

\vspace{5pt}
\noindent\textbf{Interactivity.} Modern webpages enable user interaction via hyperlinks, menus, and search boxes, whereas images are inherently static. To introduce interactivity, \cite{bhardwaj2018driveshaft} proposes \textit{click maps}, which store \texttt{<x,y>} coordinates of interactive elements. We adopt this approach, allowing \tool to notify the server (via SMS, if available) when a user clicks on a coordinate, retrieving the corresponding page if it is not already cached. Given \tool's potentially slow network conditions--seconds for uplink and minutes for downlink--we limit interactivity to hyperlinks.

\vspace{5pt}
\noindent\textbf{Image Format.} Unlike \cite{bhardwaj2018driveshaft}, which requires lossless PNG for crowdsourced screenshot merging, \tool utilizes \textit{WebP}~\cite{webp}, a modern format offering superior compression. Webpages are captured as WebP images at 10\% quality, significantly reducing file size while maintaining readability. Images are 320 pixels wide and up to 10,000 pixels tall, enabling users to \textit{scroll} with minimal data overhead. To accommodate different screen sizes, images are resized using a scaling factor (\texttt{screen width / 320}), ensuring accurate click map coordinates.

\vspace{5pt}
\label{sec:modulation}
\noindent\textbf{Modulation.} We use the \texttt{Quiet} library to modulate \tool files as waveform audio files (WAV). Inspired by ``audible-7k-channel'', we created a new modulation profile that uses Orthogonal Frequency-Division Multiplexing (OFDM)---a multi-carrier modulation technique that divides the available spectrum into multiple orthogonal narrow-band signals called sub-carriers. Our profile uses 92 sub-carriers, with a center frequency of 9.2~KHz, achieving a rate of 10~kbps.

\subsubsection{Pushing}
\label{sec:prefetching}
During low-usage periods, typically at night, \tool \textit{pushes}~\cite{suh2007push} popular webpages to its clients to ensure faster response times during peak hours. For each webpage, \tool also pushes some internal links allowing users to seamlessly interact with a webpage without delay. However, pushing every hyperlink on a webpage would overwhelm the system’s limited transmission bandwidth. To address this, we use a prioritization metric to rank internal links based on their importance:
\[
score = 0.68 \cdot w \cdot h - 0.32 \cdot y
\]
where \texttt{w}, \texttt{h}, and \texttt{y} denote the width, height, and vertical distance from the top of the page, respectively. We derive this metric from a Prolific user study detailed in Section~\ref{sec:benchmarking}.

When a webpage enters the Screenshot Queue, \tool computes this score for each hyperlink and selects the top three ranked links. These are added to a separate idle queue that is only activated when the server has no active transmissions. As \tool primarily targets informative sites like news and blogs, this metric prioritizes pages with top headlines, larger images, and prominent font sizes, ensuring that key content is already available when users attempt to follow links. Additionally, it penalizes links that are vertically farther from the top of the page.

\begin{figure}[t]
    \centering
    \includegraphics[width=1\linewidth, trim={10 10 10 10}]{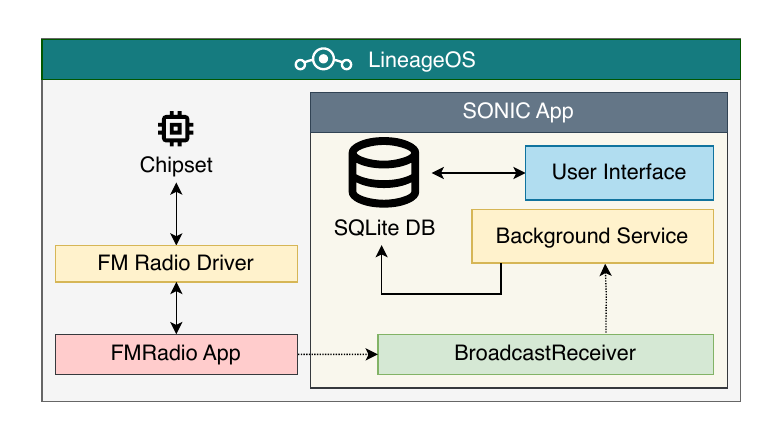}
    \caption{\tool client architecture.}
    \label{fig:sonic-client}
    \vspace{-0.2in}
\end{figure}

\subsection{\tool Client}
\label{sec:sonic-client}
At a minimum, \tool users require a smartphone with a built-in FM radio receiver, serving as the \textit{downlink}, along with a wired earphone to act as an antenna. Additionally, users who wish to send requests, such as retrieving a webpage or interacting with the LLM, need access to an SMS service for the \textit{uplink}. On the software side, \tool operates as a user-space application on a modified version of Android (see  Figure~\ref{fig:sonic-client}). We detail the \tool client in the following. 

\vspace{5pt}
\noindent\textbf{OS Integration.} The \tool client relies on FM radio hardware to receive data transmissions, but modern Android devices do not expose FM chip access to third-party applications. Default FM radio apps are shipped as system apps, integrated into the ROM and signed with privileged keys that allow hardware-level access. Apps like \tool are unable to access FM audio without rooting the device and allowing superuser access, which is unrealistic for adoption and raises significant security concerns. As such, enabling FM-based decoding requires changes at the operating system level.

We build a proof-of-concept implementation based on LineageOS~\cite{LineageOS2024}, a widely supported open-source Android distribution. We modify its default FM radio app to: 1) allow tuning the FM chip to a specific frequency, and 2) forward decoded audio streams to other apps like \tool without needing root or elevated permissions. This is achieved by implementing a \texttt{BroadcastReceiver}~\cite{android_broadcast_receiver} in the \tool app and a matching sender in the default FM radio app that transmits raw audio buffers in real time. This change allows \tool to passively listen to FM broadcasts and decode data as it arrives. We verified this approach by flashing our customized LineageOS build onto devices that use Qualcomm Snapdragon chipsets, including Xiaomi Redmi Go. We observe that as long as FM radio drivers are available, this approach can be extended to any other chipset. Manufacturers would implement this change when building the stock operating system for devices they ship with FM radio support (see Section~\ref{sec:discussions}).

\begin{figure}[t]
    \centering
    \setlength{\fboxsep}{0pt}%
    \fbox{\includegraphics[width=0.325\linewidth]{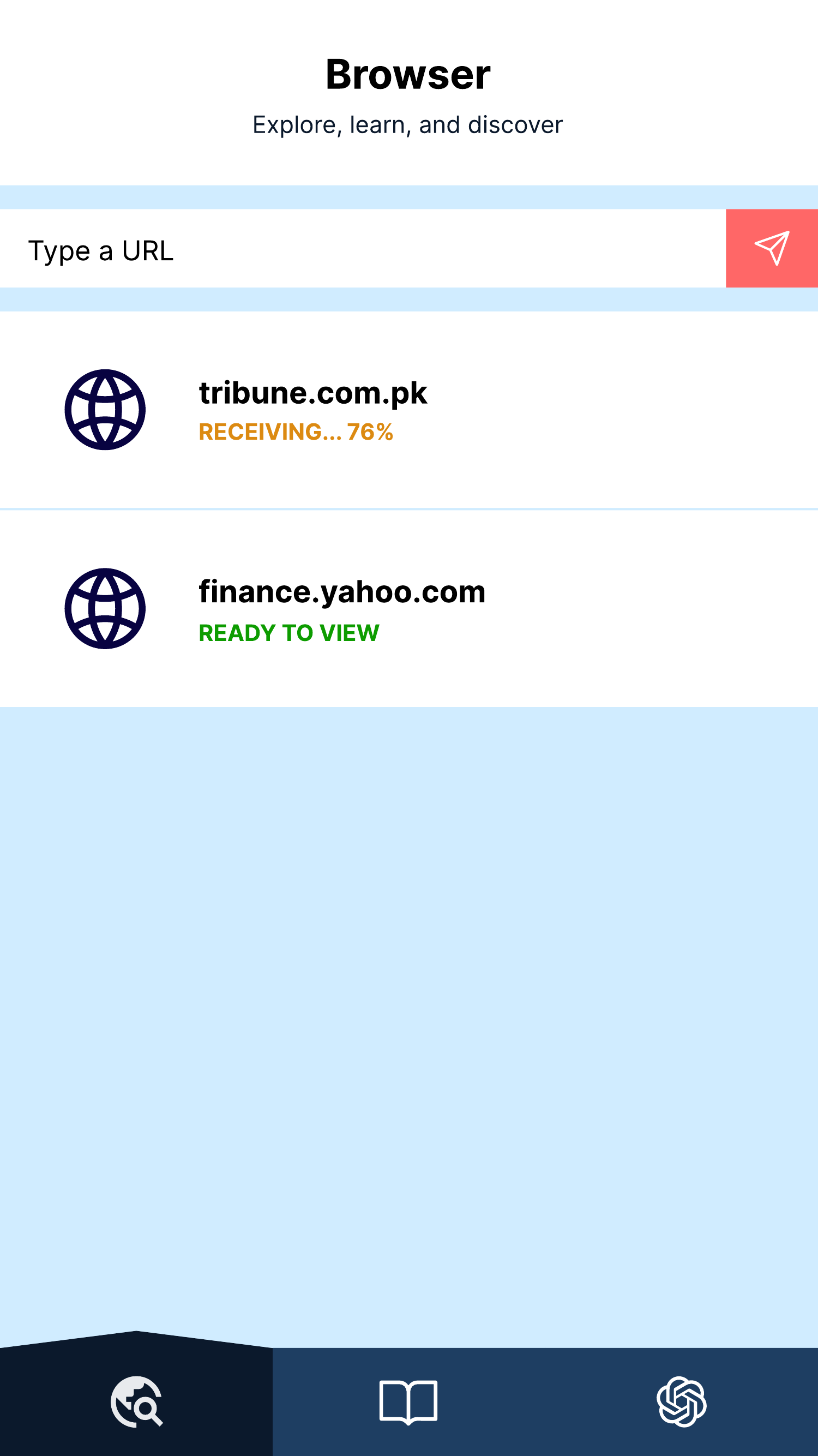}}\hspace{1pt}%
    \fbox{\includegraphics[width=0.325\linewidth]{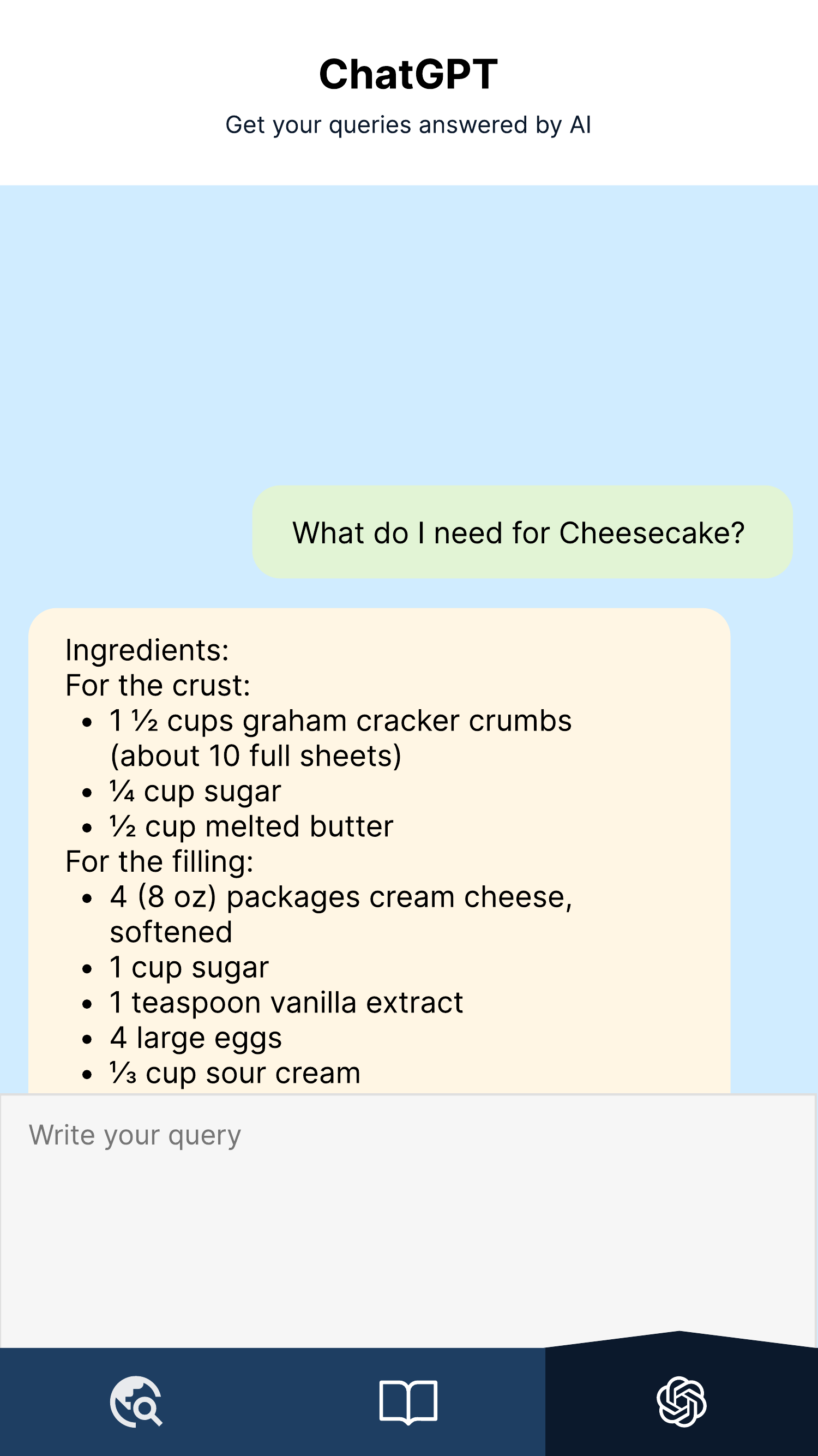}}\hspace{1pt}%
    \fbox{\includegraphics[width=0.325\linewidth]{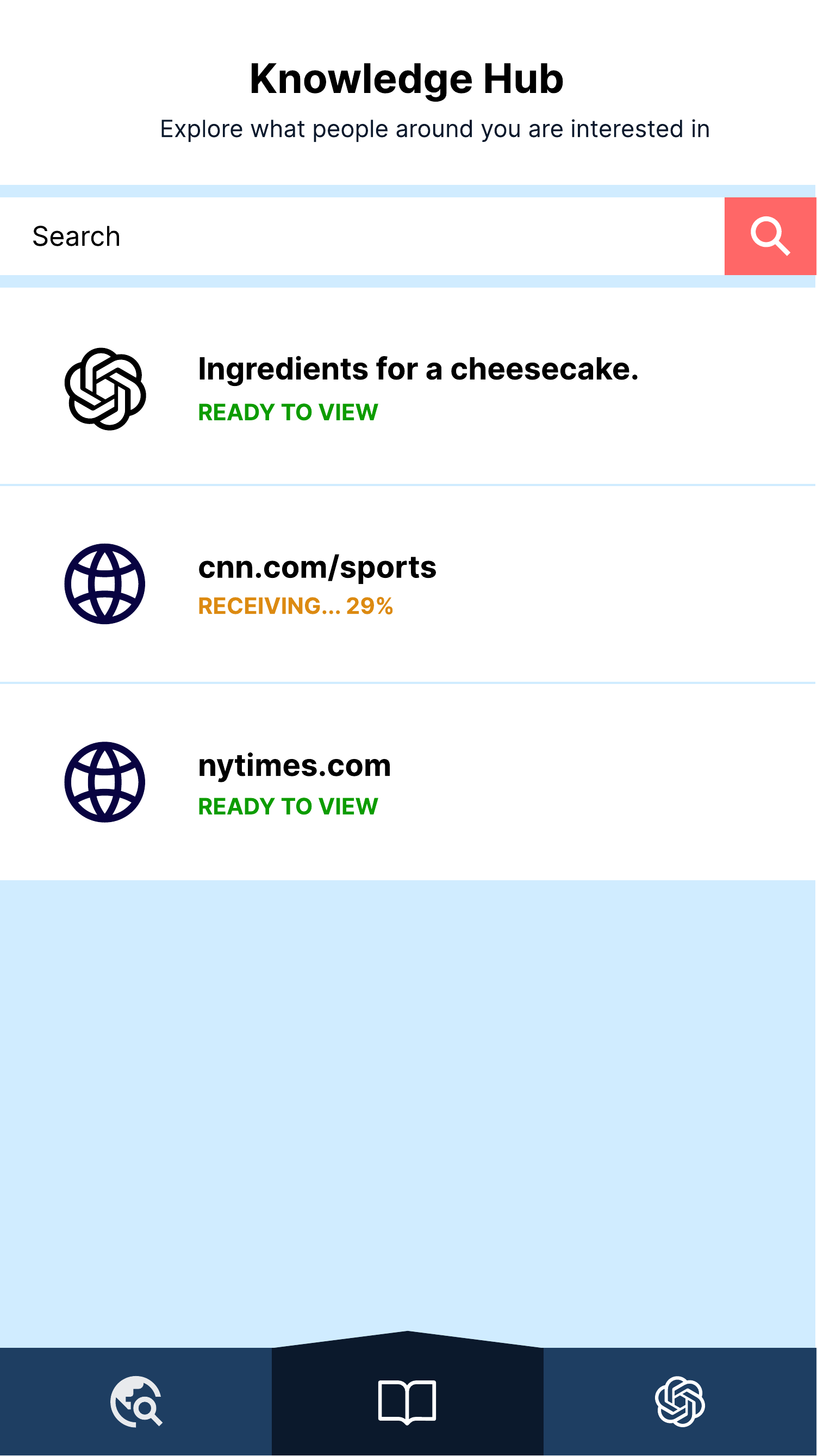}}
    \caption{\tool's user interface.}
    \label{fig:sonic-mobile-app}
    \vspace{-0.2in}
\end{figure}

\vspace{5pt}
\noindent\textbf{User Interface.} Figure~\ref{fig:sonic-mobile-app} shows the \tool app’s user interface (UI), which consists of three sections: Browser, ChatGPT and Knowledge Hub. The Browser section mimics modern web browsers. It features a search bar at the top, allowing users to request URLs. A list of requested URLs is also displayed, and once received, they appear as ``ready to view.'' The ChatGPT section provides a chat-like interface where users can interact with ChatGPT by asking questions and receiving responses. Since FM radio operates as a broadcast system, all devices tuned to the same frequency receive the same content, even if it wasn't specifically requested by those devices. This broadcast nature of FM radio is leveraged by \tool with its Knowledge Hub section. In this section, users can access a list of webpages and ChatGPT responses that are popular within their region, allowing them to discover trending content shared by others nearby.

\vspace{5pt}
\noindent\textbf{Background Service.}
In addition to the UI, the \tool app runs a background service that performs two key functions: 1) continuously listening to FM radio audio streams at a specific frequency, and 2) decoding transmissions when a \tool-encoded signal is detected. The background service listens to bytes broadcasted by the default FM radio app using a \texttt{BroadcastReceiver}. To decode \tool-encoded transmissions, it uses a modified version of \texttt{Quiet}'s Android library~\cite{quietAndroid}, which by default decodes audio from the device's microphone. We modified the library to accept bytes retrieved from the phone's FM radio app and decode them using the same modulation profile described in Section~\ref{sec:modulation}.

This background service operates independently of the user interface and automatically stores all incoming transmissions in an SQLite database. To optimize storage, any unaccessed content is automatically deleted after one day.

\vspace{5pt}
\label{sec:error-correction}
\noindent\textbf{Error Correction.} We utilize crc32 checksums per frame to detect errors. Furthermore, an inner FEC scheme (v29) and an outer FEC scheme (rs8) are used to correct transmission errors. For completely lost frames, the \tool receiver applies nearest-neighbor pixel interpolation~\cite{nearest-neighbor-interpolation}, replacing missing pixels with the value of their adjacent left pixel given that webpage consists mostly of text read from left to right.  

In the literature, better-performing techniques exist to recover missing pixels in images~\cite{gradient-compressive-sensing, deep-learning-data-reconstruction, deep-image-inpainting}, leveraging deep neural networks to learn patterns and structures of the image or utilizing sparsity and gradients in the data to fill in the missing regions. These techniques are both memory- and CPU-intensive, far beyond what a low-end mobile device can support today. Thus, we adopt a lightweight approach proposed and benchmarked by~\cite{sonic_conext} that provides consistently high content readability scores even at a 20\% pixel loss rate. 

\subsection{Discussion}
\label{sec:discussions}

\vspace{5pt}
\noindent\textbf{Rollout.} Governments, NGOs, and other organizations can roll out \tool across a wide range of Android phones by pre-installing a modified version of the LineageOS ROM (detailed in Section~\ref{sec:sonic-client}). Moreover, smartphone manufacturers could adopt this approach natively when designing their operating systems. Manufacturers can enable support for data-over-FM use cases like \tool without compromising system security or requiring root access by bundling a modified FM radio app that exposes decoded audio streams to other applications. This would allow future devices to support FM-based services out of the box with minimal engineering overhead.

\vspace{5pt}
\noindent\textbf{Incentives and Monetization.} \tool users benefit by gaining access to a streamlined version of the Web in areas where such access is typically unavailable. For providers, one approach is to charge users directly. However, this can be difficult since users receiving content via downlink are passive, making it hard to know when or if content is being accessed. As an alternative, providers could link the service to SMS, allowing paying users to request content on demand, while keeping access free for others. Notably, FM broadcasting costs remain constant, regardless of the number of listeners.

A more promising revenue model mirrors how traditional radio stations function: expanding the audience to increase advertising revenue. \tool adds a unique offering that could draw more users, potentially enhancing ad-based profits. Furthermore, ads are no longer limited to audio—they can now include visuals embedded in the web pages.

\vspace{5pt}
\noindent\textbf{Privacy Concerns.}  At a high level, \tool resembles acceleration platforms like Google AMP~\cite{amp} and WebLight~\cite{webLight}, which modify webpage content before sending it to users. These services typically rely on access to both the URLs users request and the content they consume, which raises potential privacy concerns. While a \tool server could, in principle, gather enough information to build user profiles, it avoids this issue by using FM radio as a broadcast channel. This mode of delivery makes it impossible to identify who is receiving the content. As a result, users on the downlink side remain fully anonymous, passively receiving data initiated by others nearby, without any associated privacy risk.

\vspace{5pt}
\noindent\textbf{Limitations.} \tool does not enable access to login-restricted content, such as online banking or social media accounts. This is unfeasible for downlink-only users (\ie no SMS support); for uplink users, it would involve sharing login credentials with the \tool server which is a significant privacy risk. Moreover, because content is broadcasted, any personalized information (like account details) would be exposed to anyone within range, further compromising privacy.

Next, \tool lacks support for video which is a major part of modern web usage, \eg streaming, news, and social media. \tool's limited bandwidth makes video streaming infeasible. Instead, video content is replaced with static, non-interactive thumbnails. Likewise, advanced features driven by JavaScript or CSS are not supported, as \tool only transmits simplified, pre-rendered versions of webpages.
\section{Benchmarking}
\label{sec:benchmarking}
This section benchmarks \tool under controlled lab settings. We begin by analyzing the relationship between RSSI (Received Signal Strength Indicator) and packet loss. Next, we benchmark the impact of such losses on the user experience. We conclude evaluating the effectiveness of \tool pixel interpolation and pushing techniques.

\begin{figure}[t]
    \center
    \includegraphics[width=0.95\linewidth, clip]{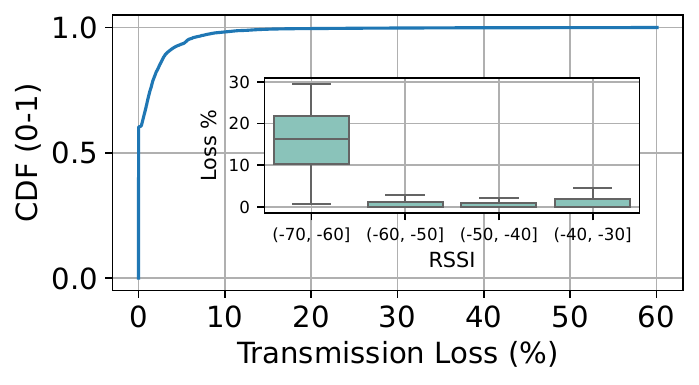}
    \caption{CDF of loss percentage under ideal conditions. Inset shows loss \% vs RSSI.}
    \label{fig:benchmarking-ideal-conditions}
\end{figure}

\vspace{5pt}
\noindent\textbf{Signal Strength and \tool Performance.} We place five Xiaomi Redmi Go phones, each with the \tool app installed, at varying distances from a 0.5~W FM transmitter to artificially create diversity in RSSI. All devices are kept fully powered meanwhile 5,000 randomly-selected webpages from the Tranco~\cite{tranco} list are broadcasted over FM at a frequency of 91.5~MHz, so that they are concurrently received by the testing devices while emulating varying RSSI. Figure~\ref{fig:benchmarking-ideal-conditions} shows the CDF of loss rates across all transmissions and its inset plot presents loss percentage as a function of RSSI range. The main CDF shows that over 95\% of transmissions experienced loss rates below 10\%. Furthermore, high loss percentages (for the remaining 5\% of transmissions) are largely confined to relatively poor signal conditions, particularly in the RSSI range of –70 to –60~dBm. As RSSI improves, the loss percentage rapidly declines and stabilizes near zero.

\vspace{5pt}
\noindent\textbf{Pixel Interpolation.} \cite{sonic_conext} evaluates the impact of visual loss and pixel interpolation on perceived content clarity and text readability using feedback from 151 Pakistani university students across 50 test webpages. Their results show that even at a 20\% pixel loss rate -- which is rare in \tool as shown in Figure~\ref{fig:benchmarking-ideal-conditions} -- users reported a median content clarity score of 7 out of 10, indicating a generally clear understanding of the page. While text readability was more affected, it remained acceptable at a loss rate of 20\%. Rating distributions from this study are provided in Appendix~\ref{appendix:pixel-interpolation}.

\vspace{5pt}
\noindent\textbf{Pushing Metric.} We conduct a user study on Prolific~\cite{prolific2025} to evaluate the likelihood of hyperlink clicks based on visual features of each link. Specifically, we examine the area covered by the link in a webpage screenshot (width $\cdot$ height) and its vertical position on the page (y-position). We randomly sample 100 webpages from the Tranco list. For each page, we generate mobile screenshots along with the bounding-box coordinates (x, y, w, h) of every hyperlink. We then create an interactive webpage where participants are asked to click on the link they would ``naturally'' choose to visit next.

\begin{figure}[t]
    \center
    \includegraphics[width=\linewidth, clip]{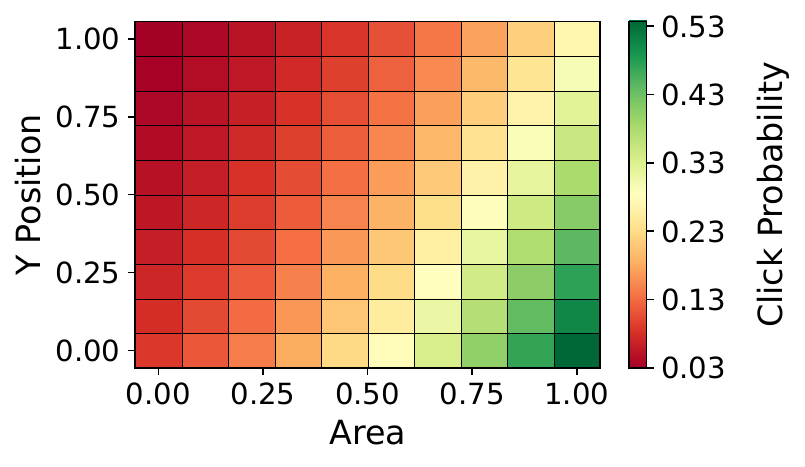}
    \caption{Likelihood of clicking on a hyperlink in a webpage given its area and y position. Data from 1,000 page interactions from 100 users across 100 pages.}
    \label{fig:prefetching-heatmap}
    \vspace{-0.2in}
\end{figure}

A total of 100 participants take part in the study, each interacting with 10 different pages. The pages are distributed so that each webpage is evaluated by ten users, resulting in 1,000 total page interactions. Using the data collected from our study, we train a logistic regression model to learn the relative importance of a hyperlink’s area and vertical position in predicting click likelihood. The model fits a weighted linear combination of these features to estimate the probability of a link being clicked. The resulting scoring function is:
\[
score = 0.68 \cdot w \cdot h - 0.32 \cdot y
\]

where \texttt{w}, \texttt{h}, and \texttt{y} denote the width, height, and vertical distance from the top of the page. The negative weight on y-position reflects that links appearing closer to the top (\ie with lower y-values) are more likely to be clicked. We use this scoring function as our prioritization metric for pushing.

Figure~\ref{fig:prefetching-heatmap} shows the likelihood of clicking a hyperlink based on its area and vertical position on the page. The heatmap reveals that links with larger areas and located closer to the top (\ie lower y-values) are more likely to be clicked. This trend is visible in the gradient transition from red (low click probability) to green (high click probability), moving from the bottom-left to the top-right of the plot. Although the maximum observed click probability is only 0.53---indicating that clicks are far from guaranteed---the relatively higher likelihood still offers a useful signal for prioritization. Since the server remains underutilized during idle periods, pushing these links—even at moderate click probabilities—can improve user experience with minimal additional cost.
\section{Deployment}
\label{sec:deployment}
This section outlines \tool deployment at a live FM radio station in Cameroon. We selected this location given its low internet penetration rate comparable to low-income regions (58.1\% of Cameroon's population is offline as of 2025~\cite{datareportal2025cameroon}).

\vspace{5pt}
\noindent\textbf{Methodology.} We start by signing an agreement with an FM radio station \textit{[name and exact location redacted for anonymity]} in Cameroon to allow the \tool server to transmit content from 10PM to 5AM daily for six weeks. This overnight window was the only available airtime, as the station's daytime schedule was reserved for regular programming. Such opportunistic use of off-peak radio hours represents the most feasible adoption path for \tool in the near term. In the future, we envision dedicated FM channels operating full-time for data broadcasting.

We recruited 30 Cameroonians to experiment with \tool during this period, \ie request webpages and ask questions to ChatGPT. Study participants were given a Xiaomi Redmi Go phone (featuring Qualcomm Snapdragon 425 processor and 1~GB RAM) flashed with the modified version of LineageOS, and \tool app pre-installed. To send requests, each phone had a SIM card with an unlimited SMS bundle. 

\begin{figure}[t]
    \centering
    \setlength{\fboxsep}{0pt}%
    \fbox{\includegraphics[width=.48\linewidth, clip]{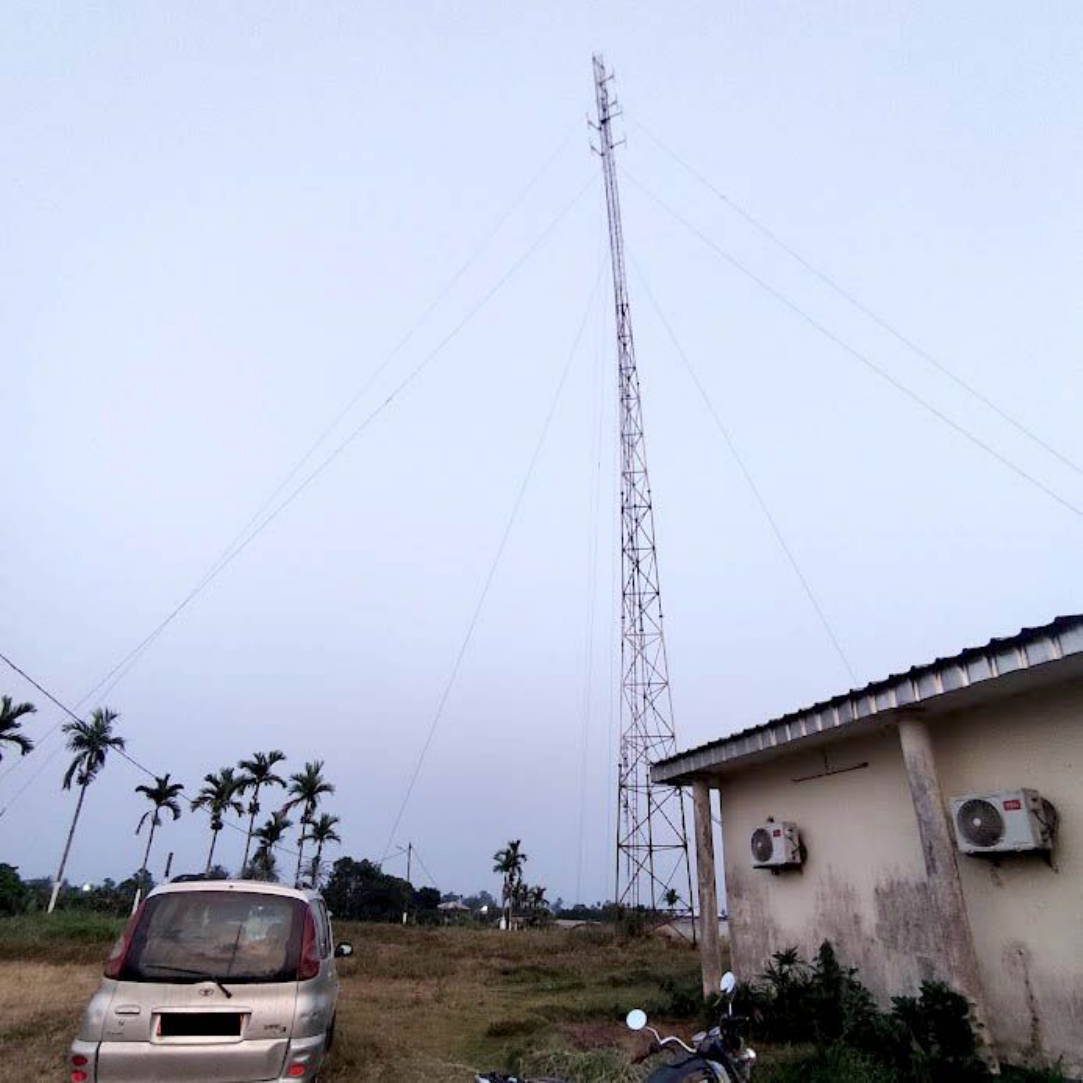}}\hspace{1pt}%
    \fbox{\includegraphics[width=0.48\linewidth]{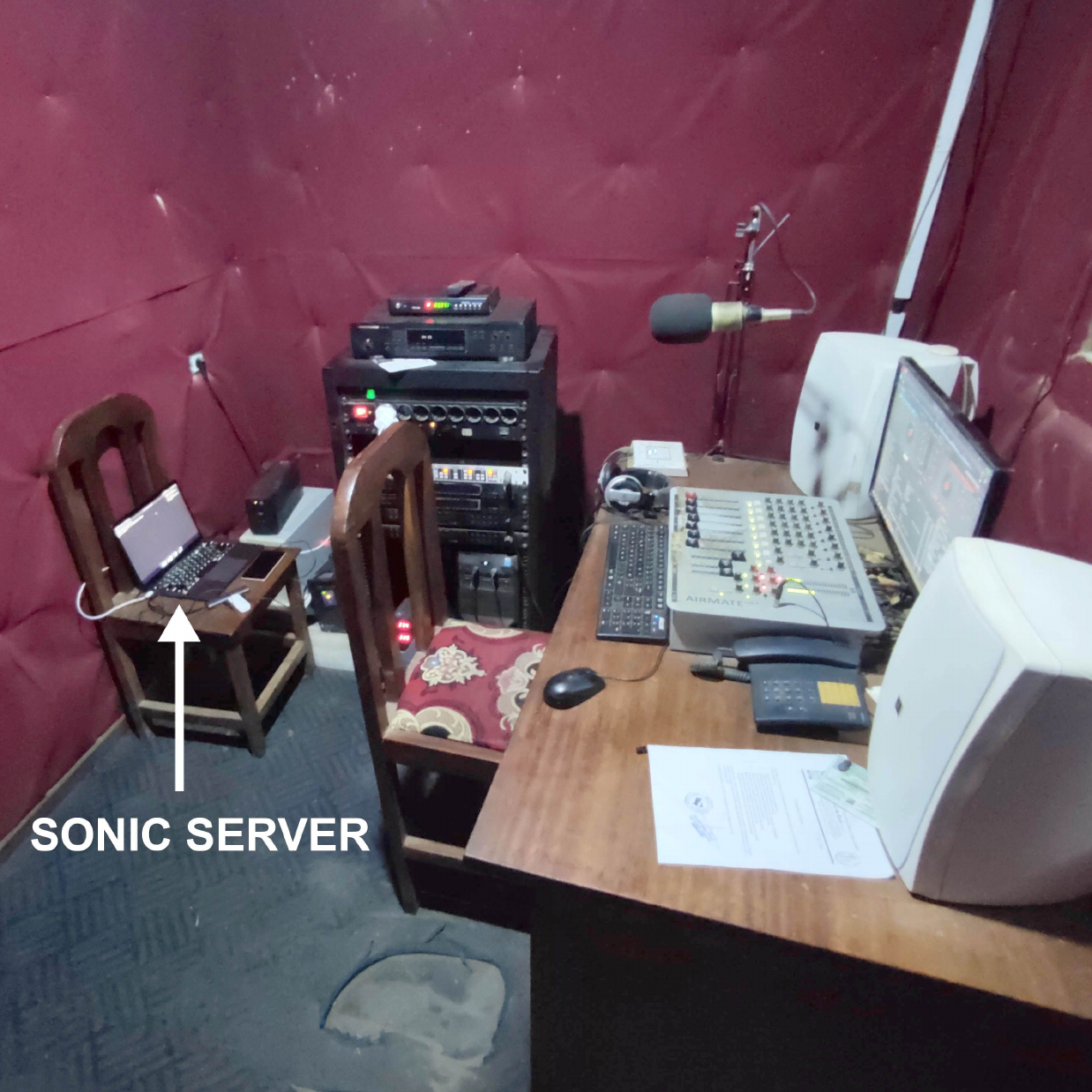}}
    \caption{FM radio station in Cameroon, and \tool server located inside the station.}
    \label{fig:server-at-the-radio}
\end{figure}

As shown in Figure~\ref{fig:server-at-the-radio}, the \tool server was set up at the FM radio station using a MacBook Air with 8 GB of RAM, running the \tool Docker container. We used Huawei's E8372h-320 LTE/4G USB Mobile WiFi Dongle~\cite{huaweiE8372h517} to interface with the SIM card via \texttt{huawei-lte-api}~\cite{huaweiLTEAPI} and receive incoming SMS messages. The \tool app was programmed to send SMS messages to the number associated with the SIM card used by the dongle. For internet access, we used a mobile data subscription from MTN Cameroon (see Section~\ref{sec:motivation} for details on plan and connection quality).

\begin{figure}[t]
    \centering
    \includegraphics[width=\linewidth, clip]{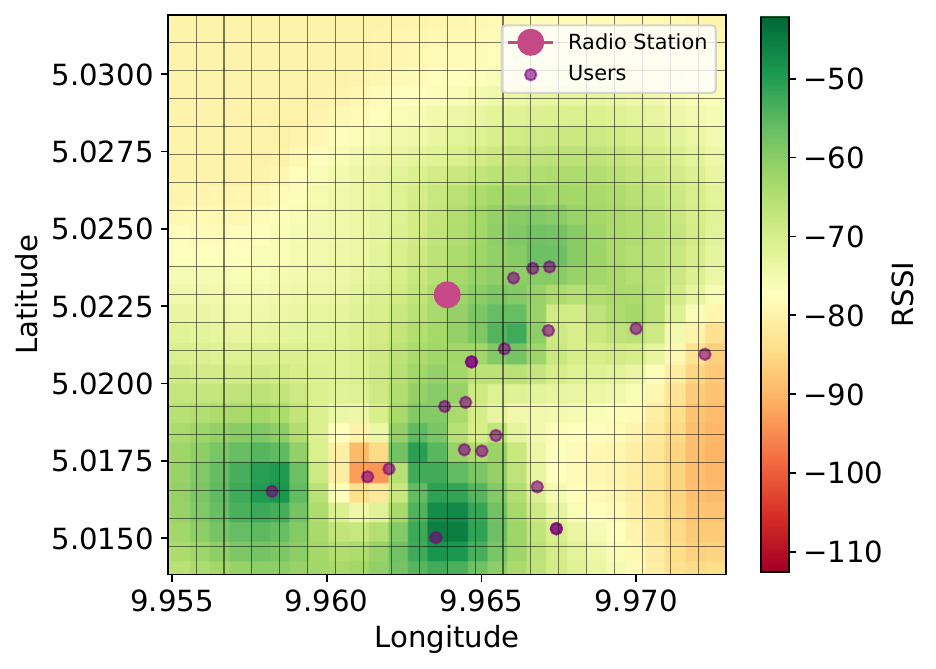}
    \caption{Location map of the 30 \tool users with their interpolated RSSI values. Each cell in the grid is 100 meters in scale. Note that some users are overlapping.}
    \label{fig:rssi-map}
\end{figure}

\vspace{5pt}
\noindent\textbf{Data Collection.} Table~\ref{table:deployment-summary} outlines our deployment. A total of 30 participants were recruited, divided into two sequential batches of 15 participants each. The study spanned 6 weeks in total, with each batch participating for 3 weeks. As an incentive to the participants, we offered USD 2 per person per day. Before the study, participants signed a consent form and were allowed to withdraw from the study at any time. The participants were then given an introduction on how to use the \tool app; further, an institutional review board (IRB) approval was granted to conduct the study. The authors who conducted the study are CITI~\cite{citi} certified. No sensitive or personal information of the participants was collected, except for their name and phone number to contact them and disburse the incentive money at the end of the experiment.

Study participants were allowed to make up to 10 requests per day--this included both webpage URL requests and GPT queries. Participants were allowed to make requests at anytime during the day; however, responses were transmitted during the transmission window of 7 hours (10~PM to 5~AM). Over 3 weeks, participants made 1,737 URL requests and 2,936 GPT queries in total. The median number of requests per user was 160, with 96 GPT queries and 64 URL requests.

\begin{table}[!t]
\small
\begin{tabularx}{\linewidth}{r X}
    \hline
    \textbf{Property} & \textbf{Description} \\ \hline
    Location                  & Cameroon \\
    Participants              & 30 \\
    Number of batches         & 2 \\
    Participants per batch    & 15 \\
    Duration of study         & 6 weeks \\
    Duration per batch        & 3 weeks \\
    Transmission window       & 10~PM to 5~AM daily \\
    Daily request quota       & 10 (GPT + URL) \\
    Total URL requests        & 1,737 \\
    Total GPT requests        & 2,936 \\
    Median requests per user  & Total: 160, GPT: 96, URL: 64\\
                              & (in 3 weeks) \\ \hline
    \end{tabularx}
\caption{Summary of \tool deployment.}
\label{table:deployment-summary}
\vspace{-0.2in}
\end{table}

\vspace{5pt}
\noindent\textbf{Deployment Challenges.} We encountered several challenges during \tool deployment. Initially, airport security confiscated 10 mobile phones intended for participants, significantly reducing the number of devices available for deployment. Only 15 phones ultimately reached Cameroon, forcing us to conduct the experiment in two separate batches. Securing reliable internet connectivity for the \tool server also proved challenging: even the best available mobile internet plan from MTN was unstable (as discussed in Section~\ref{sec:motivation}), occasionally unavailable for entire days, and affected by significant latency. Compounding these issues, the village where we deployed frequently experienced electricity outages—lasting up to 8 hours and often overlapping with the transmission window—completely disrupting FM radio transmissions. During these outages, although users continued to request content via SMS, they were unable to receive any responses. 

Operational issues further complicated the deployment. In the first batch, the \tool app sent acknowledgment messages (ACKs) for all received transmissions, enabled by unlimited SMS bundles purchased for each participant. However, the resulting high volume of SMS traffic quickly raised suspicion with the mobile operator, leading to the blocking of all deployed SIM cards. Consequently, we had to disable the ACK mechanism, leaving us without real-time operational feedback or heartbeat signals from participants' phones. Additionally, the success of FM radio transmissions critically depended on the radio station staff accurately switching to \tool broadcasts at exactly 10~PM each night. This switch-over, however, was inconsistent, resulting in multiple days without any transmissions. We also discovered that the transmitter’s output volume needed to be set to 100\% to achieve better range and improve transmission quality, but maintaining this setting consistently proved challenging for the radio staff as well. Finally, the effectiveness of the \tool system relied heavily on participants regularly charging their phones and keeping earphones connected at all times, as the earphones served as antennas for receiving broadcasts. Collectively, these logistical and operational hurdles made it difficult to maintain ideal conditions for \tool deployment.

\section{Results}
\label{sec:results}
In this section, we present our analysis from \tool deployment at a live FM radio station in Cameroon. 

\vspace{5pt}
\noindent\textbf{RSSI and Loss Analysis.} Figure~\ref{fig:rssi-map} shows the spatial distribution of RSSI measurements across a 100m-resolution grid around the FM radio station. Using ordinary kriging~\cite{kriging}, we interpolate user-collected GPS-tagged signal data to generate a continuous RSSI map. While the radio station is centrally located, the strongest signal regions are notably offset to the northeast, with two additional users registering high RSSI values at distances of approximately 900–950~m (one directly to the south and another to the southwest). Contrary to the expected radial decay in signal strength with distance, these observations demonstrate the influence of antenna height and placement similar to what was observed by~\cite{nepal2017}. The northeastward bias in signal strength likely results from the antenna’s physical orientation or directional configuration, while the isolated strong-signal detections at longer distances suggest favorable line-of-sight conditions.

\begin{figure}[t]
    \centering
    \includegraphics[width=\linewidth, clip]{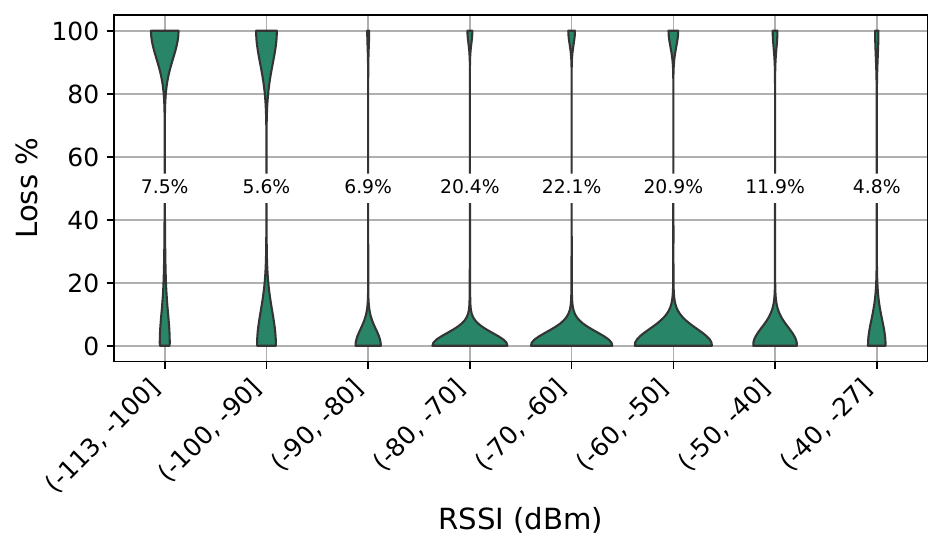}
    \caption{Loss \% as a function of RSSI range. Violin's width is proportional to the percentage of total transmissions occurred in the RSSI range.
    }
    \label{fig:rssi-loss}
\end{figure}

\begin{figure}[t]
    \centering
    \includegraphics[width=.9\linewidth, clip]{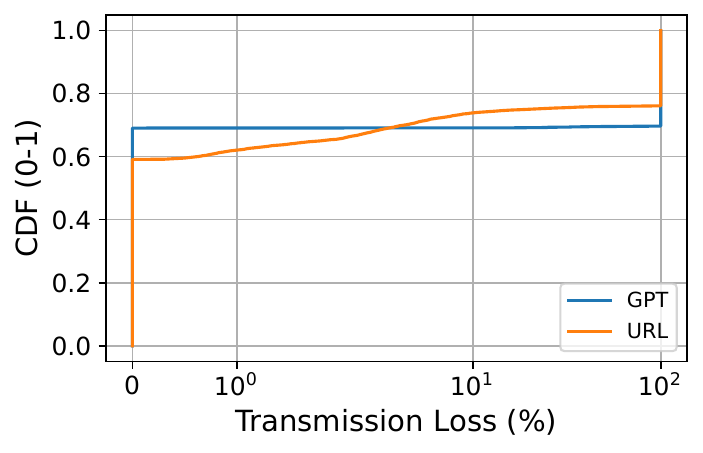}
    \caption{CDF of loss \% for GPT and URL transmissions.}
    \label{fig:cdf-loss}
\end{figure}

\begin{figure}[t]
    \centering
    \includegraphics[width=\linewidth, clip]{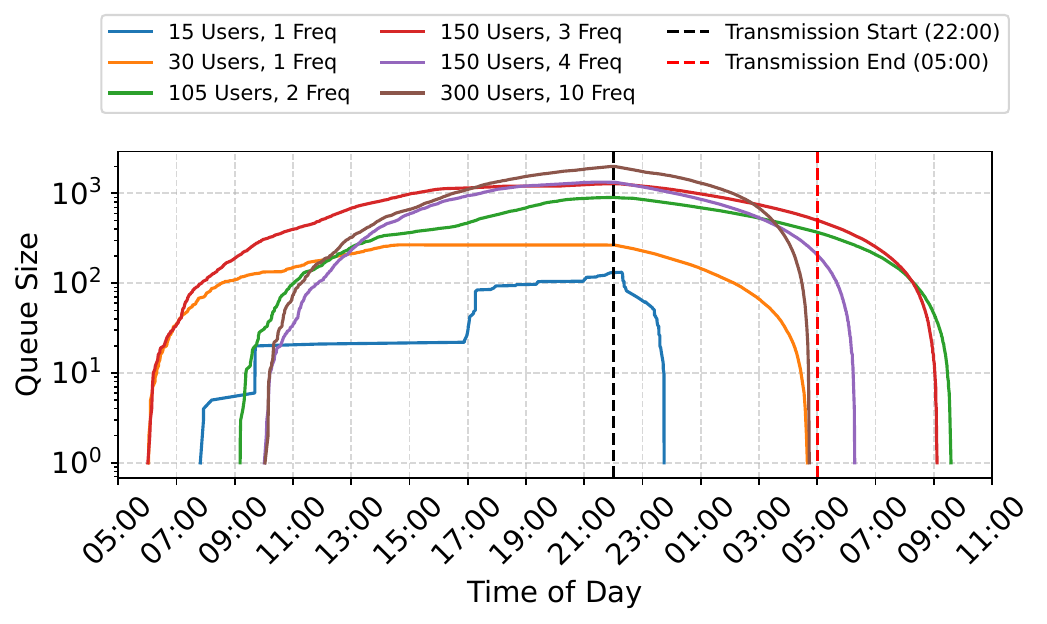}
    \caption{Queue size for \textit{actual} (15 users, 1 frequency), and \textit{simulated} (30-300 users, 1-10 frequencies) loads assuming the busiest day of \tool deployment.}
    \label{fig:scalability}
    \vspace{-0.2in}
\end{figure}

\begin{figure*}[!htb]
  \centering
    \includegraphics[width=\textwidth, clip, trim={10pt, 0, 0, 0}]{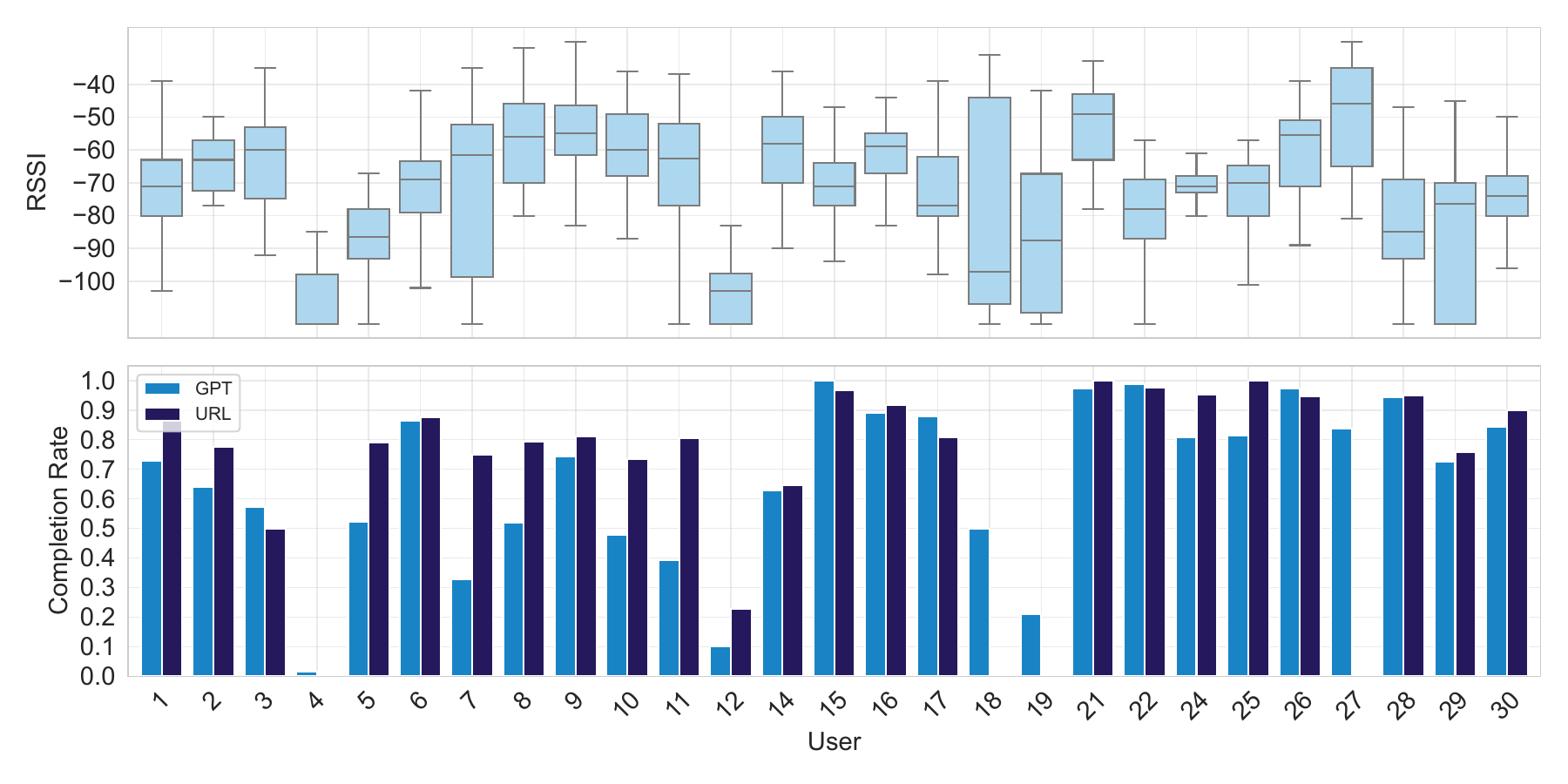}
    \caption{RSSI distribution per user with their request completion rates. Users 13, 20, and 23 were removed as their devices appeared faulty; they neither recorded any RSSI measurements nor received any transmissions.}
    \label{fig:rssi-and-completion-rate-per-user}
\end{figure*}

Figure~\ref{fig:rssi-loss} shows transmission loss percentages across RSSI ranges, with each violin’s width representing the proportion of total transmissions in that range (\eg 22.1\% of all transmissions occurred between -70 and -60~dBm). We observe a clear trend: better signal strength (\ie higher RSSI) is associated with reduced transmission loss. In poor signal conditions below -90~dBm, the loss rate is frequently near 100\%. In contrast, for stronger signals in the -80 to -50~dBm range (accounting for above 60\% of total transmissions), data points are more concentrated below 20\%, and instances of 100\% loss are rare. This confirms, in the wild, the observations from our benchmarking under controlled settings (see Figure~\ref{fig:benchmarking-ideal-conditions}). 

To investigate per user performance, Figure~\ref{fig:rssi-and-completion-rate-per-user} shows boxplots of RSSI values per participant alongside their completion rates for both GPT and URL requests. Users with stronger median RSSI values (\eg above –70~dBm) consistently achieve high completion rates across both content types. For instance, users 15–17, 21–28, and 30 report more than 80\% completion rates, indicating that strong and stable signal conditions are sufficient for reliable content delivery. In contrast, users with lower and unstable signal quality, particularly those with median RSSI fluctuating below –90~dBm (\eg users 4, 12, and 19), show drastically reduced completion rates, often below 30\%.

Due to their smaller sizes, GPT responses have shorter broadcast duration than webpages. However, they are more susceptible to failures from partial frame loss. A single 500-byte frame drop can disrupt an entire GPT message, whereas similar loss in a webpage screenshot has limited impact (see Figure~\ref{fig:pixel-interpolation}). As a result, webpages achieve higher completion rates. This can be observed for users 7, 8, 10, and 11, where URL completions are consistently higher than GPT under similar signal conditions. This trend is also reflected in Figure~\ref{fig:cdf-loss}, which shows the CDF of loss rates observed for the two types of content---GPT responses and URL transmissions. URL transmissions (orange line) appear more loss-resilient than GPT (blue line) due to their shorter duration and stricter tolerance to partial frame loss.

\vspace{5pt}
\noindent\textbf{Scalability.} Figure~\ref{fig:scalability} shows the evolution of the transmission queue over the course of a day. We select the busiest day to understand transmission trends within the transmission window (marked by the vertical dashed lines at 22:00 and 05:00). The blue trace represents the real deployment of \tool, with 15 users receiving content on a single frequency. Although each user requested 10 pages, on average, the queue size peaked at 103 items, indicating that roughly 30\% of transmissions were served from the cache. The sharp spike at 09:30 corresponds to a scheduled ``push'' of pre-selected news pages during an otherwise idle period. All other curves represent FCFS (First Come First Serve) queue simulations, scaling this baseline traffic to heavier loads. For example, with 30 users on a single FM frequency (orange), the queue peaks at around 200 items but is still fully transmitted within the 7-hour transmission window. This suggests that one frequency can support up to 30 active users with similar queuing patterns. When capacity increases to two frequencies and 105 users (green), the peak backlog rises to 800 items, half of which could not be transmitted before 05:00. Similar patterns are observed for 150 users with three and four available frequencies, respectively. However, a configuration of 300 users with 10 available frequencies appears sufficient to fully transmit the queue within the transmission window.

\begin{figure}[t]
    \centering
    \includegraphics[width=\linewidth, clip, trim={7pt 0 0 0}]{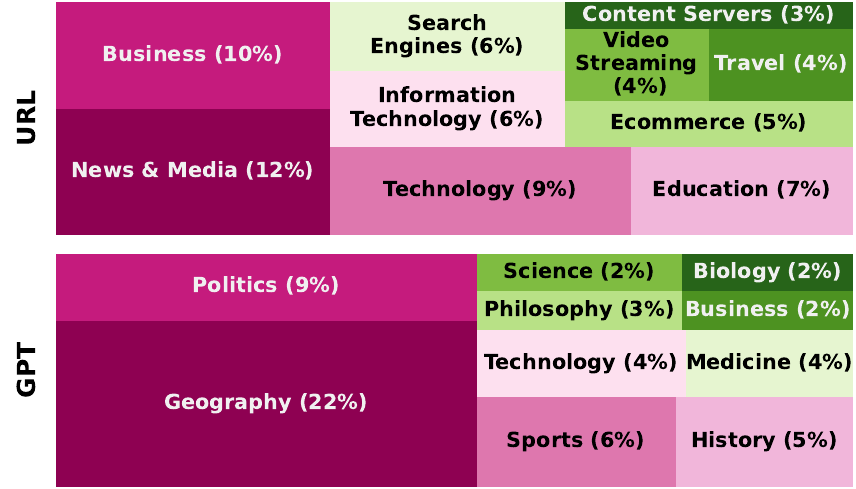}
    \caption{Top 10 categories of URL and GPT content requested by \tool users.}
    \label{fig:request-categories}
\end{figure}

\vspace{5pt}
\noindent\textbf{Content Analysis.} Figure~\ref{fig:request-categories} presents a treemap of the top 10 content categories for both GPT queries and URL requests made by \tool users. GPT queries are classified using Meta's Llama 3 model~\cite{llama3}, while domains in URL requests are categorized using Cloudflare’s Domain Intelligence API~\cite{cloudflare_domain_intel_api} following the methodology in~\cite{www_browsing_imc22}. Among GPT queries, Geography dominates with 22\% of all requests, followed by Politics (9\%), Sports (6\%), History (5\%), Technology (4\%), and Medicine (4\%). Smaller but still notable portions of queries are related to Philosophy, Science, Business, and Chemistry. For URL requests, News \& Media (12\%) and Business (10\%) are the most common, followed by Technology and Education. Users also accessed sites in categories such as Search Engines, Ecommerce, Travel, and Video Streaming. However, content from these latter categories is less likely to be useful given \tool's current limitation of only supporting page screenshots with limited (and slow) hyperlink interactivity.

\vspace{5pt}
\noindent\textbf{User Experience.} At the end of the experiment, we asked participants to complete an exit survey assessing their experience with \tool across three dimensions. Each question was rated on a 5-point likert scale, with varying response ranges depending on the aspect evaluated. For system reliability, users responded to: \textit{``How reliable was the content received through the \tool app?''} (1 = Not reliable at all, 5 = Very reliable). For UI intuitiveness, they answered: \textit{``How intuitive is the user interface of the \tool app?''} (1 = Not intuitive at all, 5 = Very intuitive). Lastly, for content relevance, users rated: \textit{``How useful was the content you received from the system?''} (1 = Not useful at all, 5 = Very useful).

\begin{figure}[t]
    \centering
    \includegraphics[width=.85\linewidth, clip]{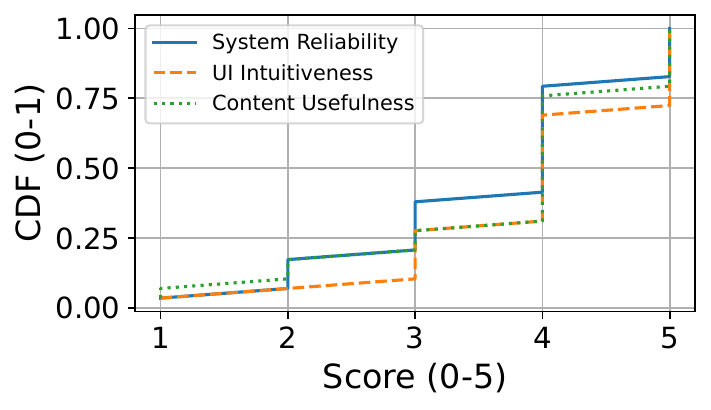}
    \caption{User experience of \tool on system reliability, UI intuitiveness, and content usefulness.
    }
    \label{fig:end-user-experience}
\end{figure}

Figure~\ref{fig:end-user-experience} shows the CDF of responses to the survey questions. Overall, user feedback was positive: 72\% of participants rated the \tool app as ``intuitive'' and the content as ``useful'' (with scores of 4 or 5). In contrast, system reliability received relatively lower ratings, with only 62\% of users selecting 4 or 5. This is expected, as some participants experienced fluctuating or low RSSI values, leading to reduced request completion rates, as discussed earlier in Figure~\ref{fig:rssi-and-completion-rate-per-user}.

\section{Related Work}
\label{sec:related}
Improving internet accessibility in developing regions is crucial for delivering essential services such as education and healthcare. However, challenges like unreliable hardware, limited cellular coverage, high data costs, and increasingly complex webpages hinder connectivity. Existing efforts focus on web simplification and new access technologies tailored for these regions. \tool combines\textit{ web simplification} with \textit{data-over-sound} transmission to enable connectivity in rural and remote areas. In the following, we discuss related works in both research areas. 

\vspace{5pt}
\noindent\textbf{Web Simplification.}  
Prior work has explored reducing webpage complexity to improve performance under limited connectivity. Habib et al.~\cite{habib10.1145/3603269.3604872} proposed a framework that dynamically adapts webpage complexity based on network conditions. Muzeel~\cite{muzeel} removes unused JavaScript, while MAML~\cite{maml} introduces a minimalist specification language omitting JavaScript and CSS. Klotski~\cite{klotski} prioritizes user-relevant content, Shandian~\cite{shandian} restructures loading via split-browser design, and Polaris~\cite{polaris} accelerates rendering using dependency graphs. Though effective under constrained bandwidth, these methods require basic internet access, which is unavailable or unaffordable in many rural areas.

\vspace{5pt}
\noindent\textbf{Data over FM.} To our knowledge, \tool is the first system to leverage FM radio as a means to broadcast internet connectivity in rural regions. The process of encoding webpages as sound is inspired by several research papers~\cite{ka2016near,lee2015chirp,santagati2016software,roy2017backdoor,bai2020batcomm} and open source tools~\cite{AudioQR,GGwave,quiet} that have explored how to transmit data over sound at inaudible audio frequencies, \ie above 18kHz. The usage of FM radio for novel applications has also been explored by in previous works including RevCast~\cite{schulman2014revcast} which leverages the broadcast nature of FM radio for certificate revocation.\cite{bozomitu2023drivers, bozomitu2022robust} use FM radio broadcasting to disseminate warning information to drivers. In 2003, Microsoft used FM subcarrier signals to turn ordinary gadgets into smart gadgets. MSN Direct~\cite{MSNDirect} was a subscription network  which sent short text updates over DirectBand, a 67.65~kHz subcarrier leased by Microsoft from commercial radio broadcasters.
\section{Conclusion}
Despite decades of progress in global connectivity, billions of people around the world remain offline—not due to a lack of infrastructure, but because of persistent affordability barriers. Access to the internet continues to be out of reach for many, especially in resource-constrained regions where even low-cost mobile data can be prohibitively expensive. In this context, \tool introduces a novel, ultra-low-cost approach to narrowing the digital divide by leveraging FM radio—a ubiquitous, inexpensive, and underutilized medium—to deliver essential web content and large language model (LLM)-based interactions without requiring an internet connection. By combining a full-system design, seamless integration with Android FM tuners, and deployment in real-world settings such as Cameroon, we demonstrate that \tool can reliably transmit simplified web content and AI-generated responses in a way that is accessible, scalable, and resilient. Our work showcases the untapped potential of repurposing existing broadcast infrastructure to extend digital access to underserved populations, offering a practical path forward for connecting the unconnected and promoting more equitable access to knowledge and services across the globe.
\bibliographystyle{plainurl}
\bibliography{main}

\appendix
\section{Pixel Interpolation}
\label{appendix:pixel-interpolation}
\begin{figure}[t]
    \centering
    \includegraphics[width=\linewidth]{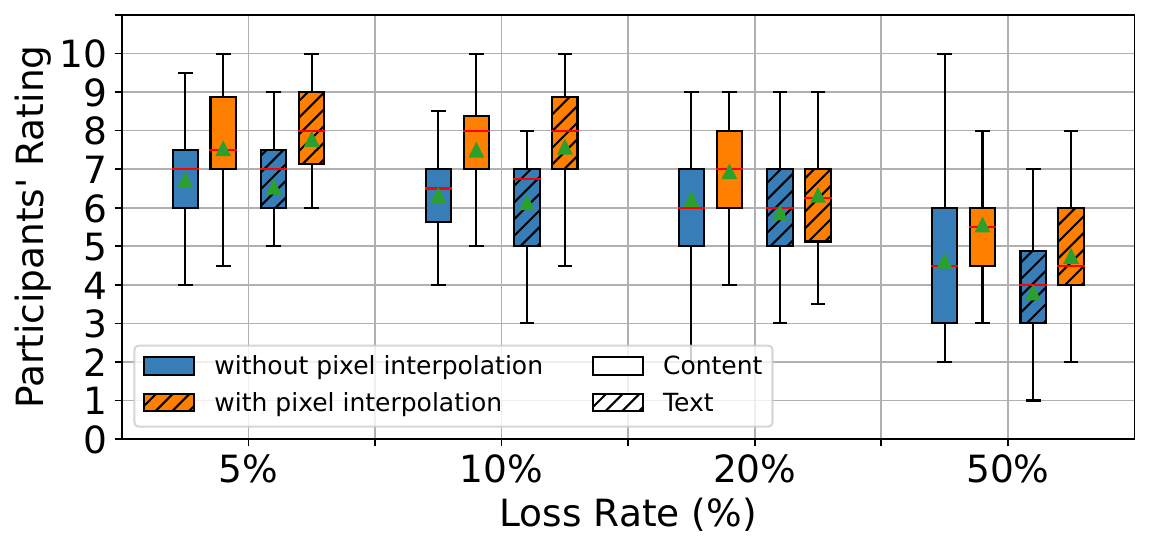}
    \caption{Distribution of user ratings (0--10) for top 50 Pakistani webpages, with/without interpolation.}
    \label{fig:pixel-interpolation}
    
\end{figure}

Figure~\ref{fig:pixel-interpolation} contains the results from a user study conducted by~\cite{sonic_conext} to benchmark the effectiveness of their pixel interpolation approach. The study simulates varying levels of visual loss on popular Pakistani webpages and evaluates their impact on perceived readability and content clarity. Screenshots of the top 50 webpages from the Tranco list were captured and processed under four levels of synthetic visual loss (5\%, 10\%, 20\%, and 50\%). Each screenshot was rendered in two variants---one with missing pixels left dark, and another corrected using nearest-neighbor pixel interpolation (detailed in Section~\ref{sec:error-correction})---yielding 400 total images.

A total of 151 university students in Pakistan participated in the study, each rating 20 randomly assigned screenshots such that each image received at least seven responses. Ratings were collected on a 0--10 Likert scale across two dimensions: (a) content clarity (\textit{How well can you understand the content in this image?}), and (b) text readability (\textit{How readable is the text in the image given the noise?}). Figure~\ref{fig:pixel-interpolation} presents median boxplots per webpage, with hatched bars representing content clarity and plain bars representing text readability. Even with a 20\% pixel loss rate, participants reported a median content clarity score of 7 out of 10, suggesting that the overall understanding of the page remained largely intact. Although text readability was more impacted, it continued to be within an acceptable range at 20\% loss.

\end{document}